\documentclass{cimento}
\usepackage{amsmath}
\usepackage{graphicx}
\usepackage[english]{babel}
\usepackage{latexsym}

\begin{document}

\title{Luminosity distance in GX cosmological models}
\author{H.G. Khachatryan\dag\ddag, G.V. Vereshchagin\ddag and G. Yegorian\dag\ddag}
\instlist{\inst{
	\dag Yerevan Physics Institute, Alikhanyan brs. st. 2, 375036 Yerevan, Armenia, \\
	\ddag ICRANet, P.le della Repubblica 10, I--65100 Pescara and \\
	ICRA, Dip. Fisica, Univ. ``La Sapienza'', P.le A. Moro 5, 00185 Rome, Italy}}

\PACSes{\PACSit{98.80.Es}{Observational cosmology}\PACSit{95.36.+x}{Dark energy}}

\maketitle

\begin{abstract}
We derive luminosity distance equation in Gurzadyan-Xue cosmological models and compared it with available supernovae and radio galaxies data sets. We found that the luminosity
distance does not depend explicitly the speed of light and the gravitation constant, and
depends only on the matter parameter (GX-invariant) and curvature.
\end{abstract}

The formula for dark energy, derived by Gurzadyan and Xue \cite{GX} defines
a relation between the speed of light $c$, the gravitational constant $G$ and
the scale factor $a$ of the Universe 
\begin{equation}
\rho _{GX}=\frac{\pi }{8}\,\frac{\hbar c}{L_{p}^{2}}\,\frac{1}{a^{2}}=\frac{%
\pi }{8}\,\frac{c^{4}}{G}\,\frac{1}{a^{2}},  \label{rhoLambda}
\end{equation}%
where $\hbar $ is the Planck's constant, $L_{p}$ is the Planck's length. The formula
does not contain any free parameters as in many approaches to the
cosmological constant problem (see e.g. \cite{Li}).

One of possible implications of this formula is variation of fundamental physical
constants with time, approach rather popular is the current literature (see e.g. \cite{Horvat}).

Based on the scaling (\ref{rhoLambda}) one may consider a set of
cosmological models \cite{Ver06}. In spite of difference of cosmological equations
in each model, some interesting similarities were found, in particular invariants, explaining underlying symmetry in the models.

In this paper we compare GX models with supernovae and radio galaxies data sets. We perform likelihood analysis and provide best-fit values of the
density parameter for all models.

Cosmological equations for GX models were derived in \cite{Ver06a} and read
\begin{eqnarray}
\dot{\mu}+3H\mu &=&-\dot{\mu}_{\Lambda }+(\mu +\mu _{\Lambda })(\frac{2\dot{c%
}}{c}-\frac{\dot{G}}{G})  \notag \\
H^{2}+\frac{kc^{2}}{a^{2}}-\frac{\Lambda }{3} &=&\frac{8\pi G}{3}\mu .
\label{MD.1}
\end{eqnarray}%
Here $\mu $ is mass density, $H$ is Hubble constant and $k=\pm 1,0$ is the
spatial curvature. The representation of the Hubble constant in the terms of
GX invariants is derived in \cite{Kha07} 
\begin{equation}
H(a)=\frac{c(a)}{a}\sqrt{\alpha\frac{a_{0}}{a}+\beta},
\label{MD.4}
\end{equation}%
where $\alpha =\frac{8\pi b_{m}^{GX}}{3a_{0}}$, $\beta =\pi ^{2}-k$.
The luminosity distance $d_{L}(z)$ as a function of redshift $z$ is defined as
\cite{Cop}%
\begin{eqnarray}
d_{L}(z) &=&a_{0}f_{k}(\kappa _{s})(1+z),\notag \\
\kappa _{s} &=&\frac{1}{a_{0}H_{0}}\int_{0}^{z}\frac{c(\acute{z})}{h(\acute{z%
})}d\acute{z},   \label{DL} \\
h(z) &=&\frac{H(z)}{H_{0}},\quad 1+z=\frac{a_{0}}{a}  \notag
\end{eqnarray}%
where $\kappa _{s}$ is normalized distance; the
subscript 0 denotes the value of each quantity today. The function $f_{k}(\kappa
_{s})$ is%
\begin{equation*}
f_{k}(x)=\left\{ 
\begin{array}{c}
\sin (x),k=1 \\ 
x,k=0 \\ 
\sinh (x),k=-1%
\end{array}%
\right.
\end{equation*}%
Using (\ref{MD.4}) and (\ref{DL}) one can find the luminosity distance and
distance modulus for GX models%
\begin{equation}
d_{L}(z)=a_{0}(1+z)f_{k}\left(\ln \left|\frac{\sqrt{\frac{\alpha }{\beta }(z+1)+1}-1}{%
\sqrt{\frac{\alpha }{\beta }+1}-1}\,\,\frac{\sqrt{\frac{\alpha }{\beta }+1}+1%
}{\sqrt{\frac{\alpha }{\beta }(z+1)+1}+1}\right|\right)  \label{DLYz}
\end{equation}
\begin{equation}
d_{M}(z)=5\log \left(\frac{d_{L}(z)}{10\,\,pc}\right).
\end{equation}%

The observation data set consists of 71 supernovae from the Supernova Legacy
Survey \cite{Astir}, 157 \textquotedblleft Gold\textquotedblright\
supernovae of \cite{Riess}, 16 \textquotedblleft Gold\textquotedblright\
high redshift supernovae from the Hubble Space Telescope (HST) \cite{Riess1}
and 20 radio galaxies of \cite{Guerra}. In total there are 264 sources with
redshifts between zero and 1.8. We have compared GX models with
observational data using standard least square technique.
\begin{equation}
\chi ^{2}=\sum\limits_{i=1}^{264}\left(\frac{d_{M}^{Obs}(z_{i})-d_{M}^{Th}(z_{i})}{%
\sigma _{i}}\right)^{2}
\end{equation}%
\begin{figure}[th]
\begin{center}
\includegraphics[width=0.80\textwidth]{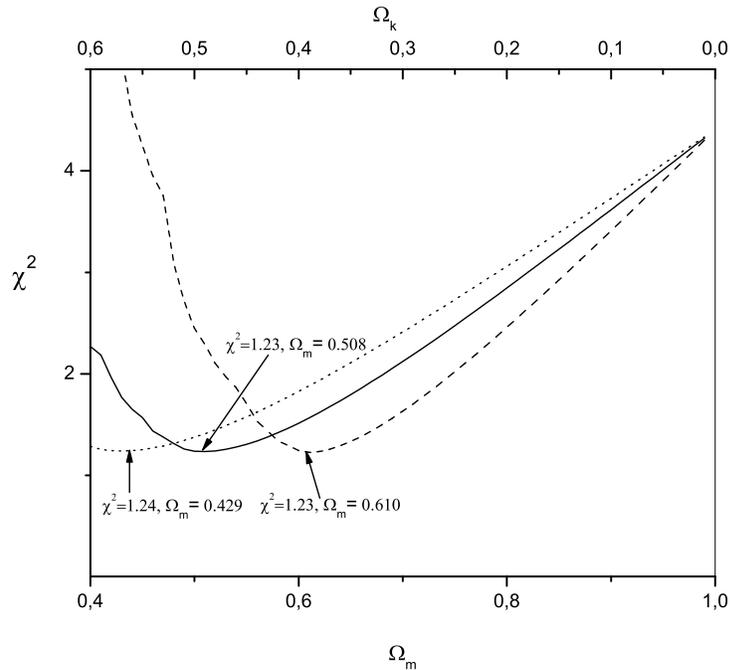}
\end{center}
\caption{ $\protect\chi ^{2}$ depending on $\Omega _{m}$ for GX models. $%
k=0,1,-1$ for thick, dash and dot lines respectively. The points of minima
are $0.43,0.51,0.61$ for $k=0,1,-1$ respectively.}
\label{fig2}
\end{figure}
here $d_{MObs}(z_{i}),d_{MTh}(z_{i}),\sigma _{i}$ are observed, theoretical
values of distance modulus and errors of data at point $z_{i}$,
respectively. Results of the fit for models are shown in fig. \ref{fig2}. As
one can see from \ref{fig2} the best fits depends on curvature $k$. The fit is better
with smaller $\Omega _{m}$.\ Since the character of solutions changes for $%
\Omega _{m}<\Omega _{sep}$ \cite{Ver06b} for GX models we took $\Omega
_{sep} $ as a best fit.

To conclude, we have derived the luminosity distance in GX models in terms of GX invariants.
It is show the important role of GX invariants as general tools to examine
the features of the cosmological models with Gurzadyan-Xue dark energy. We
performed likelihood analysis and provided best-fit values of the density
parameter for all models. We found that likelihood for all GX models
coincide. This is due to the fact that the luminosity distance depends only
on the matter density and the curvature.

\

\end{document}